\documentclass[apj,numberedappendix]{emulateapj}
\usepackage{xcolor}
\usepackage{amsmath}
\usepackage[colorlinks=true,citecolor=black,linkcolor=black,urlcolor=blue]{hyperref}
\usepackage{multirow}
\usepackage{ulem}

\newcommand{\bfx}{{\mathbf{x}}}
\newcommand{\bfk}{{\mathbf{k}}}
\newcommand{\xc}{{\bfx_\mathrm{c}}}
\newcommand{\fnoise}{{f_\mathrm{n}}}
\newcommand{\fpsf}{{f_\mathrm{p}}}
\newcommand{\dd}{{\mathrm{d}}}

\usepackage{color}
\definecolor{darkgreen}{rgb}{0.0,0.5,0.0}

\shorttitle{PSF Centroiding}
\shortauthors{Lu et al.}

\begin{document}


\title{An Accurate Centroiding Algorithm for PSF Reconstruction}

\author{Tianhuan Lu\altaffilmark{1}, Wentao Luo\altaffilmark{1},
  Jun Zhang\altaffilmark{1},Jiajun Zhang\altaffilmark{1}, Hekun Li\altaffilmark{1}, Fuyu Dong\altaffilmark{1}, Yingke Li\altaffilmark{1}, Dezi Liu\altaffilmark{2}, 
  Liping Fu\altaffilmark{3}, Guoliang Li\altaffilmark{4}, Zuhui Fan\altaffilmark{2}}

\altaffiltext{1} {Department of Physics and Astronomy, Shanghai Jiao Tong University, 
  Shanghai, 200240, China; E-mail:
  \href{mailto:njpipeorgan@sjtu.edu.cn}{njpipeorgan@sjtu.edu.cn}, 
  \href{mailto:wentao.luo82@sjtu.edu.cn}{wentao.luo82@sjtu.edu.cn}}
  
\altaffiltext{2} {Department of Astronomy, School of Physics, Peking University, 
  Beijing 100871, China}

\altaffiltext{3} {The Shanghai Key Lab for Astrophysics, Shanghai Normal University, 
  Shanghai 200234, China}

\altaffiltext{4} {Purple Mountain Observatory, Chinese Academy of Sciences, 
  Nanjing, 210000, China}

\begin{abstract}
  In this work, we present a novel centroiding method based on Fourier space Phase Fitting (FPF) for Point
  Spread Function (PSF) reconstruction. We generate two sets of simulations to test our method.
  The first set is generated by GalSim with elliptical Moffat profile and strong 
  anisotropy which shifts the center of the PSF. The second set of simulation is drawn from CFHT i band stellar imaging data. We find non-negligible anisotropy
  from CFHT stellar images, which leads to $\sim 0.08$ scatter in unit of pixels using polynomial fitting method \cite{Vakili2016}.  And we apply FPF method to 
  estimate the centroid in real space, this scatter reduces to $\sim 0.04$ in $\mathrm{SNR}=200$ CFHT-like sample.   In low SNR (50 and 100) CFHT-like samples, the background
  noise dominates the shifting of the centroid, therefore the scatter estimated from different methods are similar. We compare polynomial fitting and FPF using GalSim simulation with optical anisotropy. 
  We find that in all SNR (50, 100 and 200) samples, FPF performs better than polynomial fitting by a factor of $\sim 3$. In general, we 
  suggest that in real observations there are anisotropy which shift the centroid, and FPF method is a better way to accurately locate it.
  
\end{abstract}


\keywords{techniques: image processing; methods: data analysis}

\section{Introduction}
\label{sec_intro}
Point Spread Function (PSF) is one of the major systematics in weak lensing
measurement. It introduces both multiplicative bias and additive bias. 
There are numerous methods in literature devoted to 
correcting PSF effects
\citep{Kaiser1995, Bertin1996, Maoli2000, Rhodes2000, vanWaerbeke2001,
  Bernstein2002, Bridle2002, Refregier2003, Bacon2003, Hirata2003,
  Heymans2005, Zhang2010, Zhang2011,Bernstein2014, Zhang2015, Luo2017}.  
  Lensfit \citep{Miller2007, Miller2013, Kitching2008} applies a
Bayesian based model-fitting approach; BFD (Bayesian Fourier Domain)
method \citep{Bernstein2014} carries out Bayesian analysis in the
Fourier domain, using the distribution of unlensed galaxy moments as
a prior, and the Fourier\_Quad method developed by \citep{Zhang2010,
  Zhang2011, Zhang2015, Zhang2016} uses image moments in the Fourier Domain.

Many simulations are generated to test the accuracy of various methods, e.g. STEP (Shear TEsting Program)
\citep{Heymans2006, Massey2007}, Great08 \citep{Bridle2009}, Great 10
\citep{Kitching2010}, GREAT3 \citep{Mandelbaum2014} or Kaggle -- the
dark matter mapping competition\footnote{Supported by NASA \& the
  Royal Astronomical Society.}. Other independent softwares, such as 
SHERA \citep[hereafter M12]{Mandelbaum2012}, have also been designed
for specific surveys. But most of those simulations assume that the PSF
is perfectly known, which is not the case in reality. PSF at the position
of galaxy must be reconstructed using nearby star images.
  
In GREAT10 star challenge \citep{Kitching2013}, multiple PSF reconstruction methods has been 
tested, e.g. PSFEx \citep{Bertin2011}, PCA+Krigging \citep{LiXin2013}, Gaussianlets \citep{LiXin2013},
B-slpline \citep{Gentile2013}, Inverse Distance Weighting (IDW), Radial Basis 
Function(RBF), and Krigging \citep{ Berge2012} etc. 
Especially, \cite{Lu2016} tested various interpolation
methods to interpolate the PSF power sptectrum for Fourier\_Quad shear 
estimator \citep{Zhang2015}, which achieves $<1\%$ level accuracy in GREAT3 simulation.

 As for estimating the centroid of
stellar images, a recent work by \cite{Vakili2016} claims that simple polynomial fitting works very well and close
to saturate the Cram\'{e}r-Rao lower bound, while moment-based method 
does not deliver reliable centroid estimation. 
Our method, though, based on fitting the phase slope in Fourier space, not only
provide better centroid estimation in terms of scatter, but also automatically
shift the centroid to the center of a postage-stamp image after inverse Fourier
transformation.

We describe our method along with polynomial fitting,  in Sec.~\ref{methods}. In Sec.~\ref{Sims}, we describe the simulations to 
test our method. The results are shown in Sec.~\ref{result}. We summarize and conclude in
Sec.~\ref{summary}.

\section{Method}
\label{methods}

In this section, we describe our centeroid measurement 
method along with polynomial fitting
method with a Gaussian smooth kernel.

\subsection{Fourier space Phase Fitting (FPF)}
\label{fourier}

Given an image $I(\bfx)$ and its centroid $\mathbf{x_c}$. The Fourier transformation
of the image is simply
\begin{equation}
f(\bfk)=\frac{1}{2\pi}\iint I(\bfx)\,e^{-i\bfk\cdot \bfx}\,\dd\bfx,
\label{eqn:image-fourier}
\end{equation}
where $f(\mathbf{k})$ can be written as 
\begin{equation}
f(\bfk)=|f(\bfk)|\,e^{i\angle f(\bfk)},
\end{equation}
with $\angle (f)$ as the phase. Given that the centroid of a noise free PSF image $I(\bfx)$ is defined as 
\begin{equation}
\xc:=\frac{\iint\bfx\,I(\bfx)\,\dd\bfx}{\iint I(\bfx)\,\dd\bfx},
\label{eqn:centroid-def}
\end{equation}
we find that the 
centroid $\xc$ in real space corresponds to the slope of phase near $\bfk=0$, i.e.
\begin{equation}
\xc=-\nabla \angle f(0).
\label{eqn:centroid-relation}
\end{equation}
The proof is shown in the appendix.

In the case that the image is symmetric about the centroid, which is usually the case in the vicinity
of the centroid and the image value is real. Then according to the Fourier transformation properties
we can deduce that
\begin{eqnarray}
f(\bfk)=f(-\bfk)=f^*(\bfk).
\end{eqnarray}
As a result, $f(\bfk)$ is also a real function, meaning the imaginary part vanishes and
the phase function is zero. Any anisotropy can further introduce none
zero imaginary part, which can be reflected by the phase.


\begin{figure}
\centering
\includegraphics[width=8cm]{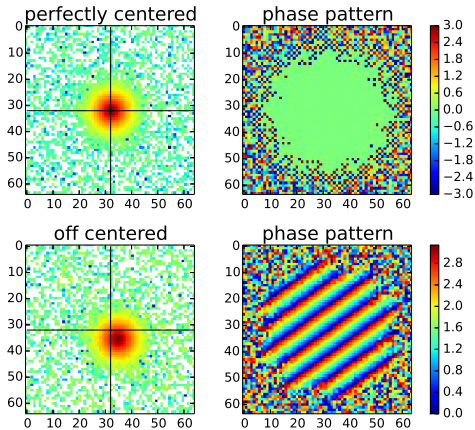}
\caption{Upper left: perfectly centered star image; Upper right: the phase at each $(k_x, k_y)$;
  Lower left: off centered star image; Lower right: phase pattern caused by off centering effect.}
  \label{fig:phase}
\end{figure}

Fig. \ref{fig:phase} gives an example of how off-center affect the phase pattern in Fourier space.
The phase will be zero if the image is perfectly centered, while there will be stripe pattern
caused by off centering effect.

After discretization, the derivative becomes a minimization problem: 

\begin{equation}
 \chi^2 = \sum\limits_\bfk{w(\bfk)\left[\angle\left(f(\bfk) e^{-i \phi(\bfk)}\right)\right]^2} 
\end{equation}
where $f(\bfk)= \fpsf(\bfk)+\fnoise(\bfk)$  is the Fourier transform of the PSF ($\fpsf(\bfk)$) 
with noise ($\fnoise(\bfk)$), and $\phi(\bfk)$ is the modeled phase pattern of $f(\bfk)$, which depends on the PSF ($f(\bfk)$), the centroid ($\xc$).
The observed image is a matrix with real values, so its Fourier transformation 
satisfies the conditions that are $f(-\bfk)=f^*(\bfk)$ and 
$\phi(-\bfk)=-\phi(\bfk)$.

For a symmetric PSF. the phase is simply related with the centroids linearly, 
\begin{equation}
\phi=k_x x_\mathrm{c}+k_y y_\mathrm{c}.
\end{equation}
Nevertheless, the higher order anisotropy that shifts centroids, should be fitted 
with higher order(here we apply 3rd order) polynomial
to capture this effect,
\begin{equation}
\phi=k_x x_\mathrm{c}+k_y y_\mathrm{c}+\alpha_1 k_x^3+\alpha_2 k_x^2 k_y+\alpha_2 k_x k_y^2+\alpha_4 k_y^3, 
\end{equation} 
where $\{\alpha_1, \alpha_2, \alpha_3, \alpha_4\}$ are free parameters. 

In the appendix, We show that the weights bearing the analytical form
\begin{equation}
w(\bfk)=
\begin{cases}
\frac{2\,|f(\bfk)|^2}{|\fnoise(\bfk)|^2}, 
&\mbox{}|\fnoise(\bfk)|\ll |\fpsf(\bfk)|\\
0, &\mbox{}|\fnoise(\bfk)|\approx|\fpsf(\bfk)|
\end{cases}
\end{equation}
can out-weight the noise while preserve as many data points as possible for the fitting.

\subsection{Polynomial fitting method}
\label{polyfit}

We follow \cite{Vakili2016} and apply a fixed Gaussian kernel 
\begin{equation}
k(x)=\frac{1}{2\pi w^2}\exp(-x^2/2w^2).
\end{equation}
where $w$ is 1.2 pixels given that the PSF FWHM is 2.8 pixels, to smooth 
the image before fitting the centroids with 2D polynomial
\begin{equation}
P(x,y)=a+bx+cy+dx^2+exy+fy^2.
\end{equation}
Only the central $3\times 3$ patch around the brightest pixels are 
used to solve the coefficients $X=\{a,b,c,d,e,f\}$. The design matrix
A can be constructed as
\begin{equation}
A=
\begin{pmatrix}
1      & x_1    & y_1    & x_1^2  & x_1 y_1 & y_1^2  \\ 
1      & x_2    & y_2    & x_2^2  & x_2 y_2 & y_2^2  \\
\vdots & \vdots & \vdots & \vdots & \vdots  & \vdots \\
1      & x_9    & y_9    & x_9^2  & x_9 y_9 & y_9^2 
\end{pmatrix}.
\end{equation}

Then the coefficients can be determined by solving
\begin{equation}
X=(A^\mathrm{T} A)^{-1} A^\mathrm{T} Z
\end{equation}
where $Z$ is the pixels of the flattened $3\times 3$ patch. 
The centroids can be then determined by those coefficients.
\begin{equation}
\begin{pmatrix}
x_c\\ y_c
\end{pmatrix}
=D^{-1}\begin{pmatrix}
-b \\ -c
\end{pmatrix},
\end{equation}
where  $D=\begin{pmatrix}
2d & e \\
e  & 2f
\end{pmatrix},$ is the curvature matrix,

\section{Simulations}
\label{Sims}
We simulated two sets of images, one uses GalSim \citep{Rowe2015} and
the other is based on Principal Components (PCs) decomposed from CFHT w2 stellar images.
The GalSim simulation provides
a set of optical effects.
The CFHT simulation is for exploring how much shifts caused by the anisotropy in real
surveys.

\subsection{GalSim stellar image}
\label{galims}
We use GalSim to simulate 3 sets of star images with different signal to noise ratio: $\mathrm{SNR}\sim50$, $\mathrm{SNR}\sim100$ and $\mathrm{SNR}\sim200$. The SNR is not strictly 50, 100 or 200 due to the fact that
we simulate the images using Exposure Time Calculator (ETC) from a uniform magnitude distribution centered at 
each SNR.

\begin{figure}
\label{fig:coma}
\centering
\includegraphics[width=0.5\textwidth]{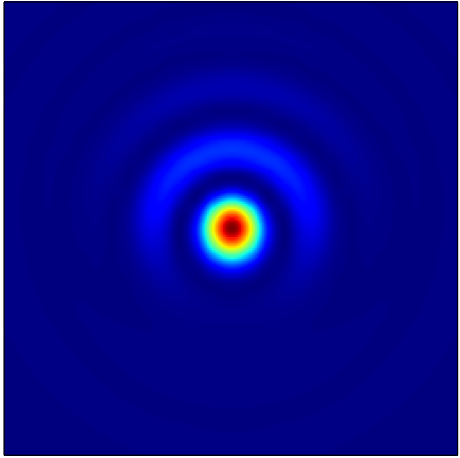}
\caption{This figure demonstrates the effect of coma with 0.07 deviation in $y$-direction from GalSim. }
\end{figure}

 We apply Moffat profile for all the stellar images, and
uniform distribution for the $r_\mathrm{d}$ and $\beta$ in 
\begin{equation}
\label{eq:moffat}
I(r)=\left(1+\frac{r^2}{r_\mathrm{d}^2}\right)^{-\beta},
\end{equation} 
where $I(r)$ is the 2D brightness distribution. 
After we generate the stellar images, we shift
the centers using two uniform distributions from $-0.5$ to 0.5 to each image as input centroid value.  Then
we convolve the images with optical anisotropy effect coma as shown in Fig. 2. We exaggerate this effect for better illustration, in fact, it is hard to be noticed by visual inspection. Finally we add Gaussian noise using
GalSim to simulate difference SNR stellar images. Noise and randomly oriented optical 
anisotropy contribute more dispersion to the centroid of the image.

\subsection{CFHT w2 stellar image}
\label{w2sims}
 In real observations, atmospheric seeing dilutes the observed objects and
introduce extra ellipticity and shift on centroid. The background sky also shifts the 
centroid randomly. For high SNR images, the centroid-shifting mechanism is dominated by
atmospheric seeing. An accurate centroid estimation method should be designed to capture
this shifting and correct for it.

In this simulation, we focus on testing this high order centroid shifting effect
based on real data from CFHTLens survey field w2, which contains 7 exposures,
with 10 minutes exposure for each. 

\begin{figure}
\centering
\includegraphics[width=8cm]{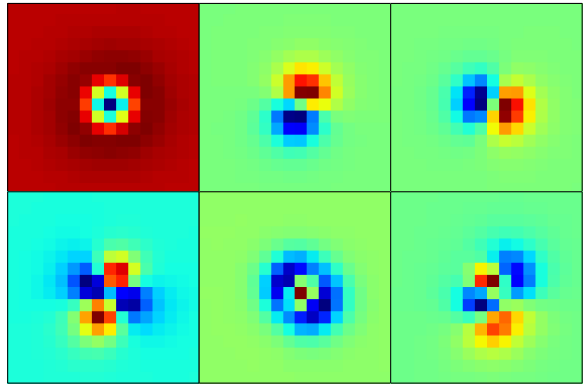}
\caption{The first six Principal Components(PCs) from CFHT w2 stellar images. }
  \label{fig:pcs}
\end{figure}

The procedure of this simulation is described as follows:
\begin{itemize}
  \item  Select all the stellar images with $\mathrm{SNR}>100$ and without saturation from CFHT w2 area, there are
   $\sim 600,000$ stars in total.
  \item  Extract components using PCA without centroiding.
  \item  Generate 10,000 stellar images using the first 16 PCs, the coefficients of the PCs are randomly drawn from the parent distribution of the original images.
  \item  Calculate the centroid using brightness weighted moments using 
   \begin{equation}
    \bfx_0=\frac{\iint
    \bfx\,I(\bfx)\,\mathrm{d}\bfx}
    {\iint I(\bfx)\,\mathrm{d}\bfx}\,.
   \end{equation} to be the reference.
   This can be considered as the real input center when there is no noise.
\end{itemize}

The first 6 PCs decomposed from CFHT w2 stellar images are shown in Fig.~\ref{fig:pcs}. We preserve the dipoles by not centroiding
the postage stamp images so that the asymmetry can be directly reconstructed by the dipoles.

\begin{figure}
\centering
\includegraphics[width=8cm]{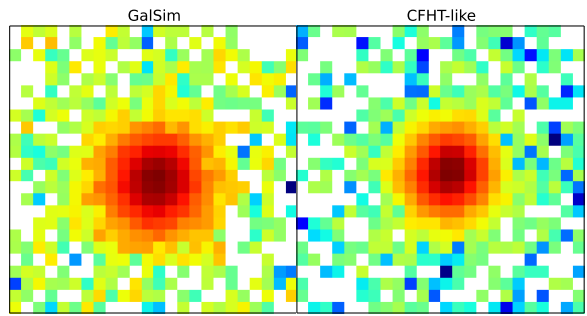}
\caption{The left panel is the PSF simulated by GalSim and the right panel is the CFHT-like PSF. }
  \label{fig:compare}
\end{figure}

We display the images from two simulations in Fig.~\ref{fig:compare}.
 The right panel is the PSF from GalSim simulation and the left one is 
CFHT-like simulation. Despite of the anisotropy effect added to GalSim PSFs, we still
can not observe its existence by visual inspection.

\section{Results}
\label{result}

\begin{figure*}
\centering
\includegraphics[width=5cm]{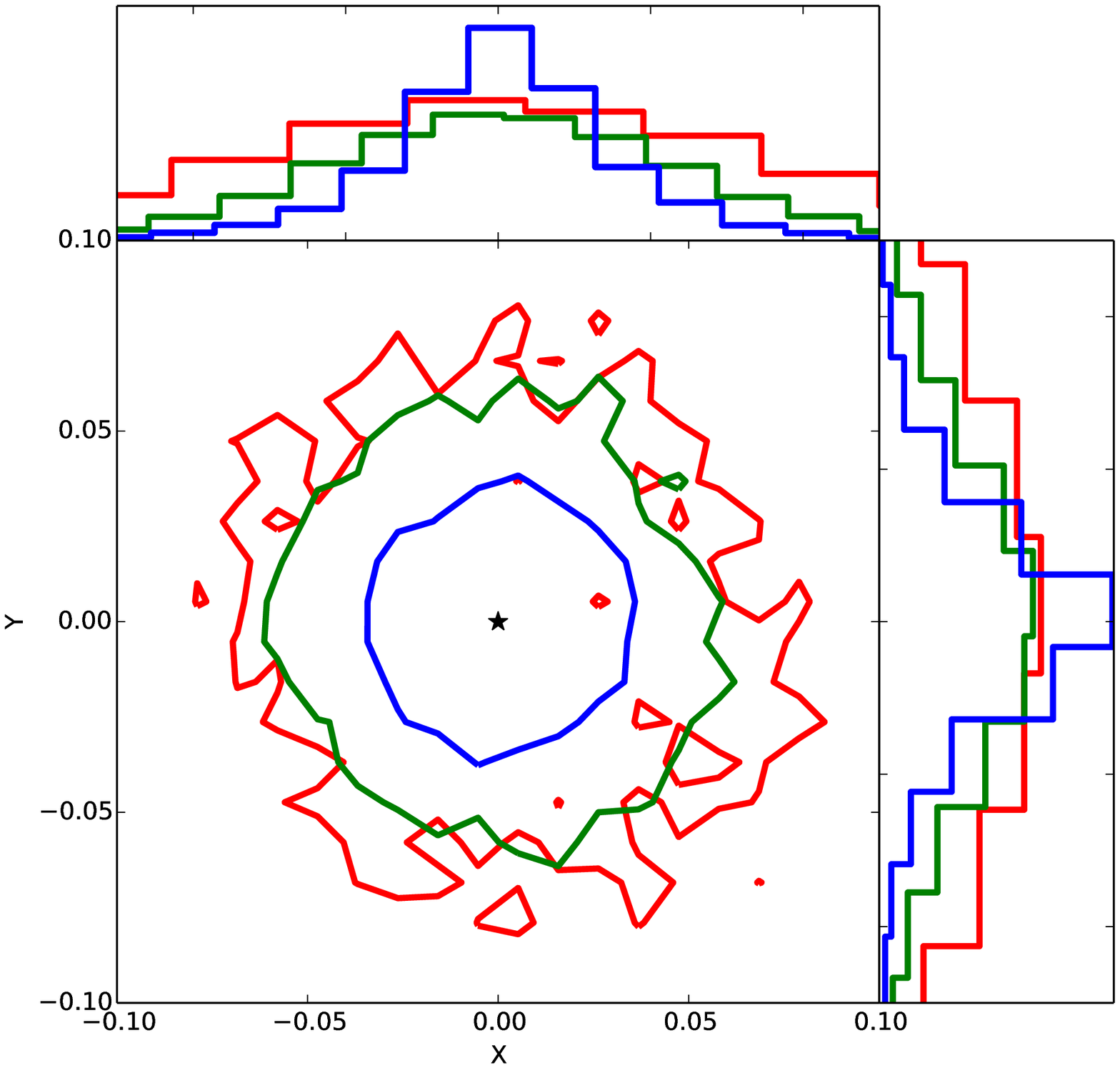}
\includegraphics[width=5cm]{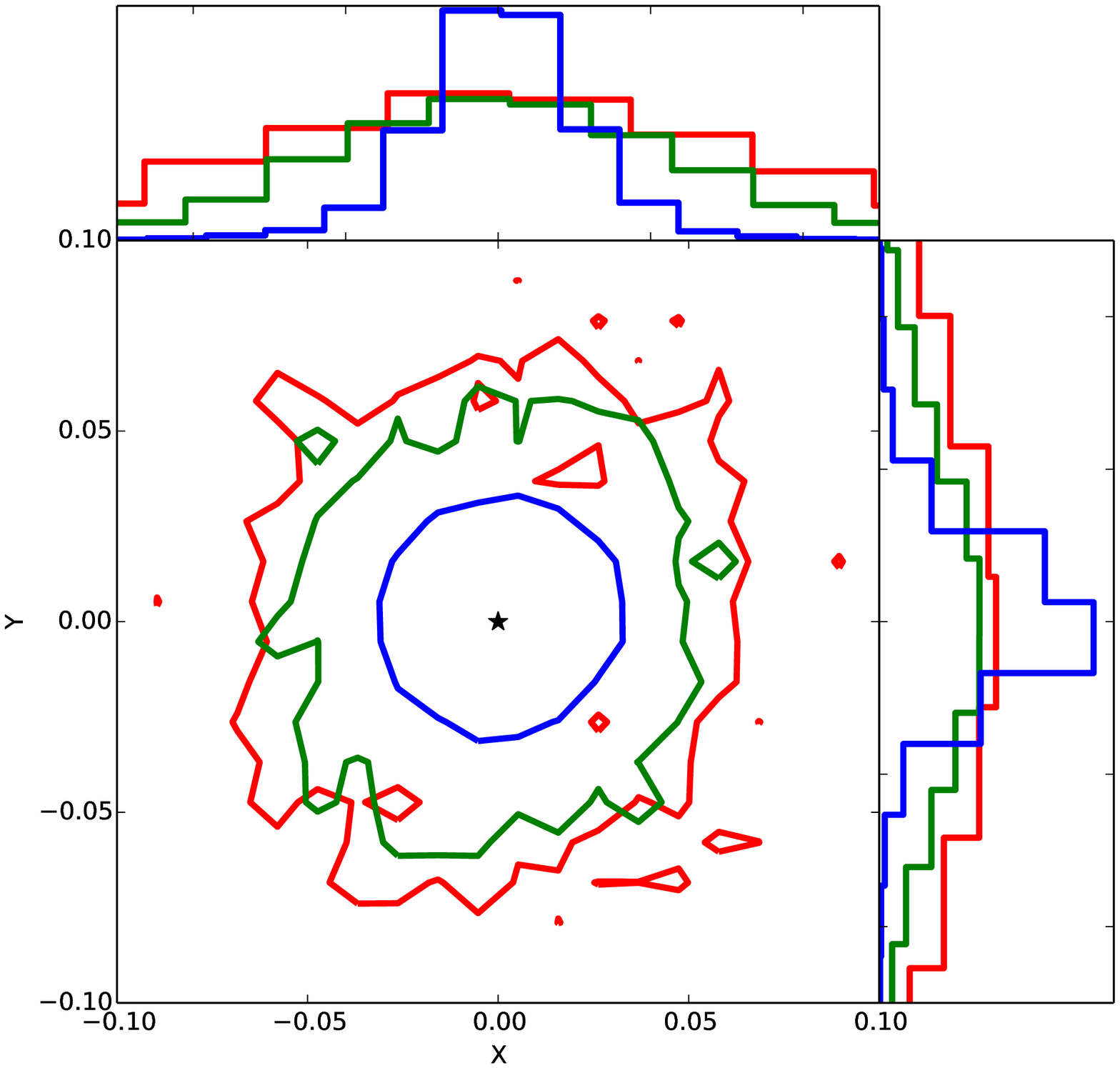}
\includegraphics[width=5cm]{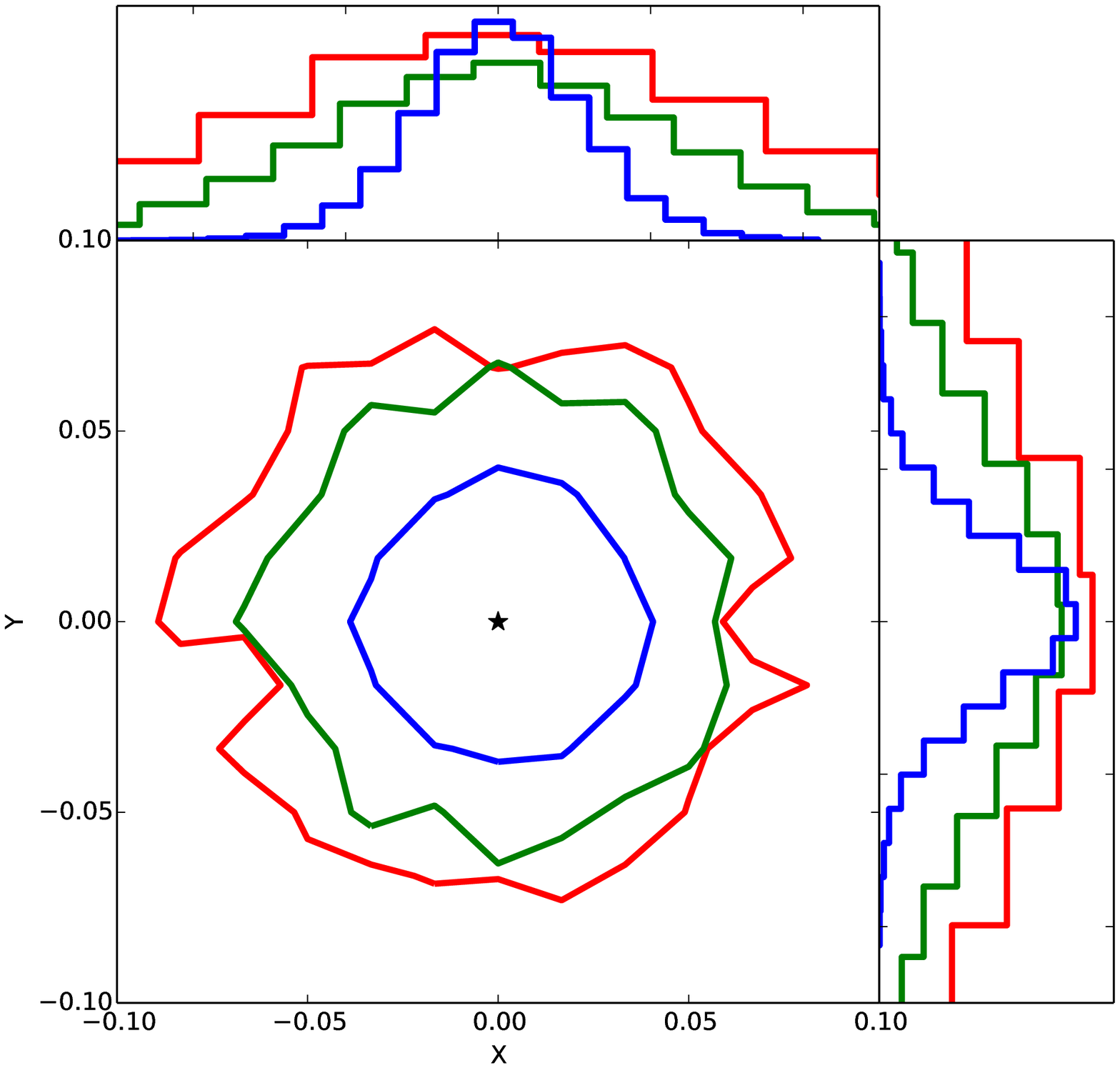} \\
\includegraphics[width=5cm]{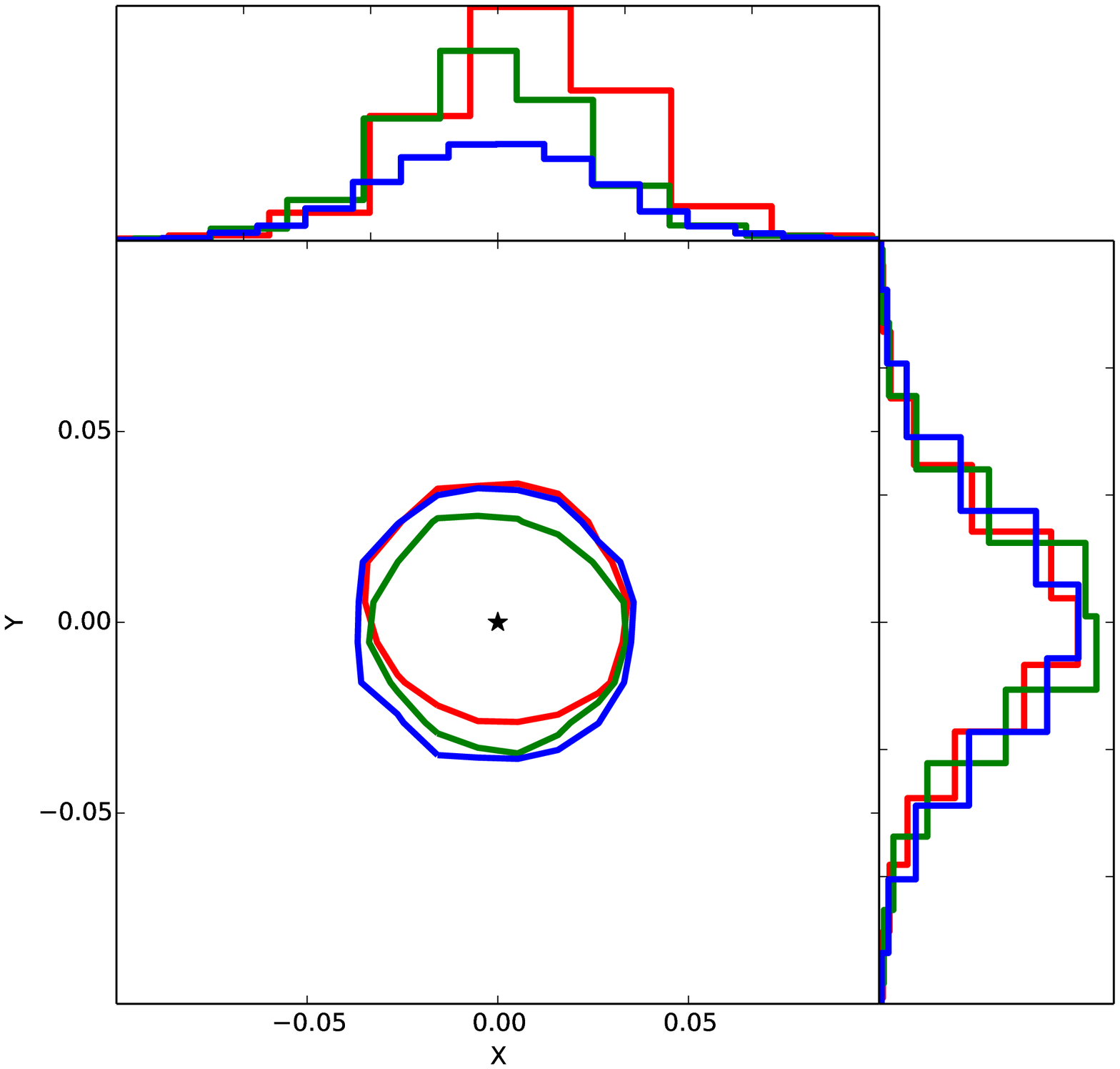}
\includegraphics[width=5cm]{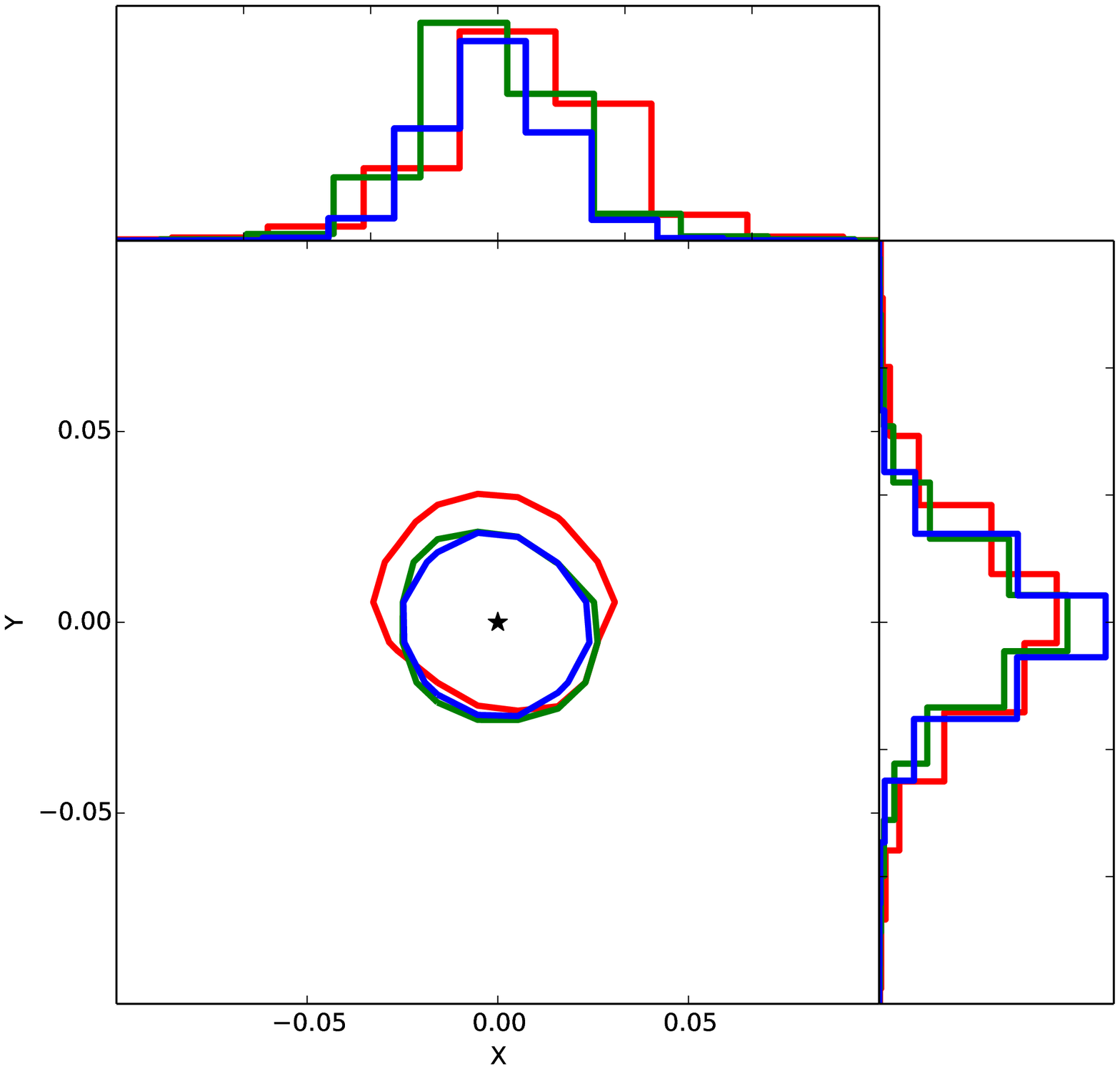}
\includegraphics[width=5cm]{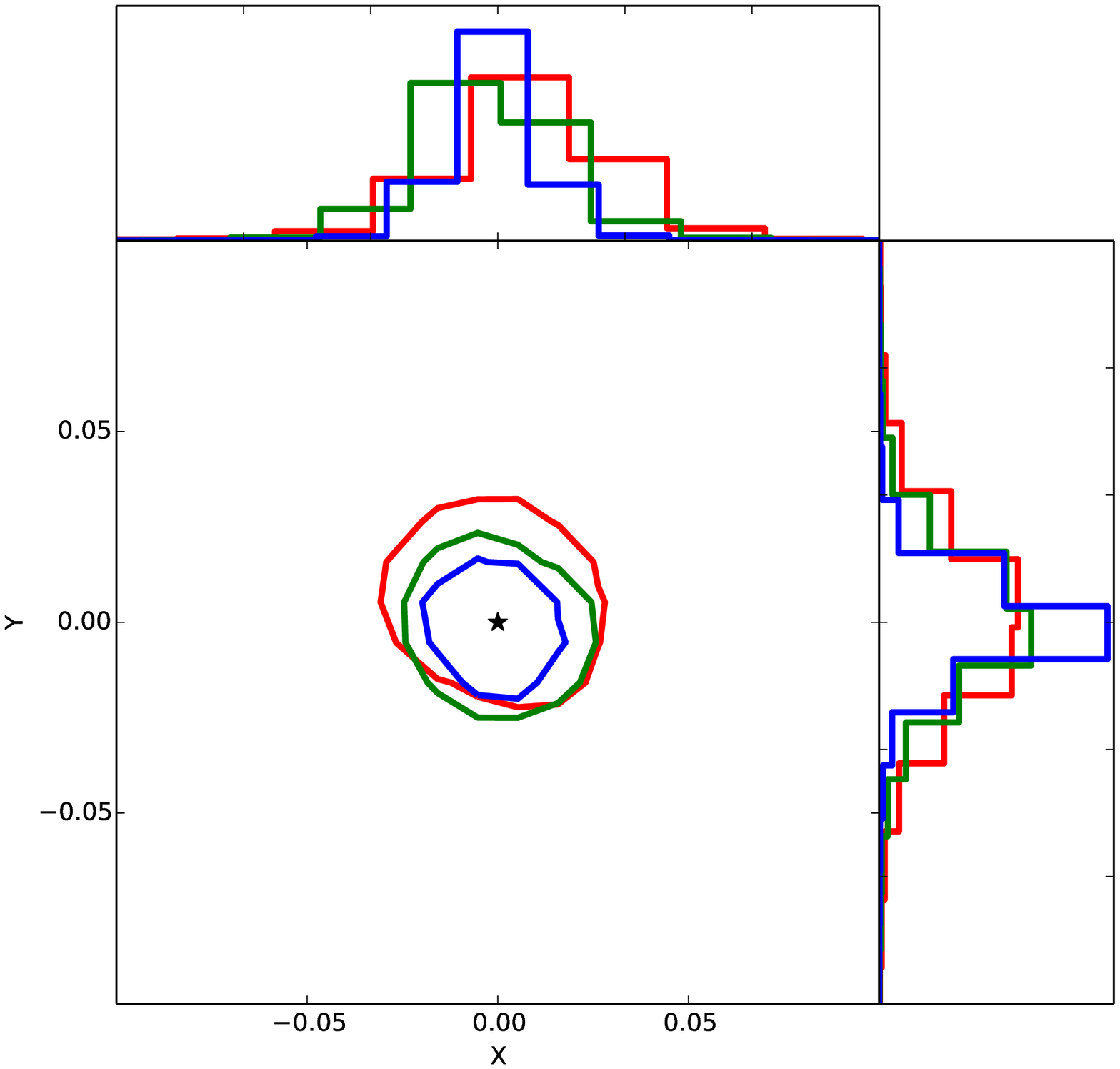}
\caption{Upper panel: Galsim simulation; Lower panel:
  CFHT w2 simulation. The red, green and blue lines denote  one sigma scatter of polynomial, 
  1st order FPF and 3rd order FPF centroiding method. The histograms are the residual distribution of centroids in x,y directions respectively.}
  \label{fig:results}
\end{figure*}

We demonstrate the performance of three centroiding
methods Fig. \ref{fig:results}, i.e. polynomial fitting, 1st order phase fitting and 3rd order phase fitting based on two simulations. The top three panels show the results from GalSim simulation, the bottom three are from CFHT-like simulation. From left to right, the $\mathrm{SNR}=50,100,200$ respectively. 

In general, 1st order FPF is already better than polynomial fitting in both simulations, and
3rd FPF is significantly improves the centroid estimation. In CFHT-like $\mathrm{SNR}=50$ simulation, 
all three methods perform similarly due to the fact that the sky background noise dominates the
scatter budget. The quantitative comparison is listed in Table.~\ref{tab:tbl-1}.

\begin{table}[h!]
\begin{center}
  \caption{\label{tab:tbl-1} The centroiding scatter comparison in unit of pixel.}
\begin{tabular}{ccccc}
\hline
\multirow{2}{*}{Simulation} & \multirow{2}{*}{SNR} & \multirow{2}{*}{Polynomial} & 1st order & 3rd order \\
 & & & FPF & FPF \\
\hline
\multirow{3}{*}{Galsim}
 & 50   & 0.3569  &  0.2350  & 0.1559 \\
 & 100  & 0.3525  &  0.2336  & 0.1127 \\
 & 200  & 0.3562  &  0.2396  & 0.1105 \\         
\hline
\multirow{3}{*}{CFHT-like}
 & 50   & 0.0877  &  0.0814  & 0.0891 \\
 & 100  & 0.0762  &  0.0601  & 0.0495 \\
 & 200  & 0.0740  &  0.0592  & 0.0377 \\   
\hline
\end{tabular}
\end{center}
\end{table}

In GalSim simulation, where the optical anisotropy dominates
the scatter budget, the scatter from 3rd order FPF is $\sim3$ times smaller
than that from polynomial fitting in all SNR branches.

In CFHT-like simulation, the performance of three methods are similar in $\mathrm{SNR}=50$ branch.
For $\mathrm{SNR}=100$ branch, which are often used in real analysis, the scatter from 3rd order FPF is $\sim1.5$ smaller than that from polynomial fitting. This difference expands to $\sim2.0$ for $\mathrm{SNR}=200$ branch. 


In the CFHT-like simulation, the performance of three methods are similar in $\mathrm{SNR}=50$ branch. As we increase SNR to 100, which is closer to those in real measurements, the scatter from 3rd order FPF is $\sim1.5$ smaller than that from polynomial fitting. This difference expands to $\sim2.0$ for $\mathrm{SNR}=200$ branch. As the effect of noise drops, 3rd order FPF has a better and better performances relative to polynomial fitting, due to its ability to accurately capture higher order anisotropy features. 

It can be further illustrated by the GalSim simulation, where higher order anisotropy dominates the scatter budget. In all SNR branches, the scatter from 3rd order FPF is $\sim3$ times smaller than that from polynomial fitting.

\section{Summary and Discussion}
\label{summary}

Centroiding is the first and important procedure for 
PSF reconstruction. An accurate PSF reconstruction
further affects shear measurement. We develop
our centroid estimation method in Fourier space -- 3rd order FPF. 


In GalSim simulation, the centroiding shift is dominated by optical anisotropy. The scatter of 3rd order FPF are smaller than polynomial fitting
method by a factor of $\sim2$ to  $\sim3$,  from $\mathrm{SNR}=50$ to $\mathrm{SNR}=200$ branch. 

In CFHT-like simulation, where the higher order anisotropy 
is much smaller than GalSim simulation, we found that for the $\mathrm{SNR}=50$
images, background noise dominate the scatter, while
for the $\mathrm{SNR}=200$, optical anisotropy play a major role for this scatter. Therefore
3rd order FPF performs similarly in $\mathrm{SNR}=50$ branch, but with half the scatter of polynomial
fitting method in $\mathrm{SNR}=200$ branch.

Therefore, we conclude that 3rd order FPF method is so far the most accurate estimation in
centroiding. The scatter caused by noise can not well corrected for any methods, but for 
scatter introduced by optical anisotropies, 3rd order FPF can capture the shift precisely. 
This is very important for weak lensing measurements.

\acknowledgements

This work was supported
by the following programs;  NSFC
(Nos. 11503064), Shanghai Natural Science Foundation, Grant
No. 15ZR1446700. JZ  is  supported  by  the  NSFC  grants  (11673016, 11433001, 11621303) and the National Key Basic Research Program
of China (2015CB857001). L.P.F. acknowledges the support from NSFC grant 11333001 and
11673018, STCSM grants 13JC1404400 and 16R1424800, SHNU grant DYL201603.

This work was also supported by the High Performance Computing
Resource in the Core Facility for Advanced Research Computing at
Shanghai Astronomical Observatory.

\appendix
\section{Centroid of a PSF image in Fourier space}

First, we calculate the derivative of $f(\bfk)$ with respect to $k_x$ at $\bfk=0$: 
\begin{eqnarray}
\nonumber\left.\frac{\partial}{\partial k_x}\right|_{\bfk=0} f(\bfk)
&=&\left.\frac{\partial}{\partial k_x}\right|_{\bfk=0} \left\{\frac{1}{2\pi}\iint I(\bfx)\,e^{-i\bfk\cdot\bfx}\,\dd\bfx\right\} \\
\nonumber&=&\left.\frac{\partial}{\partial k_x}\right|_{k_x=0}\left\{\frac{1}{2\pi}\iint I(x,y)\,e^{-i k_x x}\,\dd x\dd y\right\} \\
\nonumber&=&\frac{1}{2\pi}\iint I(x,y)\,\left.\frac{\partial}{\partial k_x}\right|_{k_x=0} e^{-i k_x x}\,\dd x\dd y \\
\nonumber&=&-\frac{i}{2\pi}\iint I(x,y)\,x\,\dd x\dd y \\
\nonumber&=&-i x_\mathrm{c}\,\frac{1}{2\pi}\iint I(x,y)\,\dd x\dd y, \\
&=&-i f(0)\,x_\mathrm{c}.
\label{eqn:dfk-1}
\end{eqnarray}
Next, we related the derivative of $f(\bfk)$ with respect to $k_x$ with that of $\angle f(\bfk)$: 
\begin{eqnarray}
\nonumber\left.\frac{\partial}{\partial k_x}\right|_{\bfk=0} f(\bfk)
&=&\left.\frac{\partial}{\partial k_x}\right|_{\bfk=0}\left\{|f(\bfk)|\,e^{i\angle f(\bfk)}\right\} \\
\nonumber&=&|f(0)|\,e^{i\angle f(0)}\left(i\left.\frac{\partial}{\partial k_x}\right|_{\bfk=0}\angle f(\bfk)\right)+\frac{\partial f(0)}{\partial k_x}\,e^{i\angle f(0)} \\
&=&i f(0)\left.\frac{\partial}{\partial k_x}\right|_{\bfk=0}\angle f(\bfk)
\label{eqn:dfk-2}
\end{eqnarray}
Compare Eqn.\eqref{eqn:dfk-1}  with Eqn.\eqref{eqn:dfk-2}, we find 
\begin{equation}
x_\mathrm{c}=-\left.\frac{\partial}{\partial k_x}\right|_{\bfk=0}\angle f(\bfk),
\end{equation}
and similarly, 
\begin{equation}
y_\mathrm{c}=-\left.\frac{\partial}{\partial k_y}\right|_{\bfk=0}\angle f(\bfk).
\end{equation}
Thus, we get 
\begin{equation}
\xc=-\nabla \angle f(0).
\end{equation}

\section{Weights in Fourier phase fitting}
Suppose the Fourier transform of the noise free PSF is $\fpsf(\bfk)$, and the noise has magnitude $|\fnoise(\bfk)|$ and random phase. Given $f=\fpsf+\fnoise$, we can deduce the scatter of $\angle f$ in terms of its variance. 

When $|\fnoise|\ll|\fpsf|$ (without loss of generality, we let $\fpsf\in\mathbb{R}$),  
\begin{eqnarray}
\nonumber\mathrm{Var}(\angle f)
&=&\left\langle(\angle f-\langle\angle f\rangle)^2\right\rangle \\
\nonumber&=&\left\langle(\angle(\fpsf+\fnoise)-\angle \fpsf)^2\right\rangle \\
\nonumber&=&\left\langle\left(\mathrm{Im}(\fnoise)\left.\nabla_\mathrm{Im}\angle f\right|_{f=\fpsf}\right)^2\right\rangle \\
\nonumber&=&\frac{1}{2}|\fnoise|^2 \left(\left.\frac{\dd f^{-1}}{\dd f}\right|_{f=\fpsf}\right)^2 \\
&=&\frac{|\fnoise|^2}{2\,|\fpsf|^2},
\end{eqnarray}
where $\nabla_\mathrm{Im}$ denotes the directional derivative along the imaginary part.

The weights should be inversely proportional to the variances. We take
\begin{equation}
w(\bfk)=\mathrm{Var}(\angle f)^{-1}=\frac{2\,|\fpsf|^2}{|\fnoise|^2}\approx\frac{2\,|f|^2}{|\fnoise|^2}.
\end{equation}

When $|\fnoise|\approx|\fpsf|$, the scatters of $\angle f$ are so large that such pixels do not provide useful information of $\angle \fpsf$. Thus we take 
\begin{equation}
w(\bfk)=0.
\end{equation}

\end{document}